\documentstyle[aps,prl,epsf,floats]{revtex}
\begin{document}
\draft
\wideabs{
\footnotesize{ \it{To appear in Phys.Rev.Lett. January 4, '99}}
\vskip 0.3cm
\title{Extraction of the Spin Glass Correlation Length}
\author{Y.G. Joh, R. Orbach, G.G. Wood}
\address{Department of Physics, University of California, Riverside,
California  92521-0101}
\author{J. Hammann and E. Vincent}
\address{Service de Physique de l'Etat Condens\'e, CEA Saclay, 91191 Gif sur
Yvette, Cedex, France}
\date{Received 18 September 1998}
\maketitle
\begin{abstract}
The peak of the spin glass relaxation rate,
$S(t)=d{{\lbrack}-M_{TRM}}(t,~{t_w})/H{\rbrack}/{d\ell n t}$, is directly
related to the typical value of the free energy barrier which
can be explored over experimental time scales. A change in
magnetic field $H$ generates an energy $E_z={N_s}{\chi_{fc}}{H^2}$
by which the barrier heights are reduced, where ${\chi_{fc}}$ is
the field cooled susceptibility {\it per
spin}, and $N_s$ is the number of correlated spins.
The shift of the peak of $S(t)$ gives $E_z$,
generating the correlation length, $\xi(t,T)$, for
$Cu:Mn~6at.\%$ and $Cd{Cr_{1.7}}{In_{0.3}}{S_4}$.  Fits to
power law dynamics, $\xi(t,T)\propto {t}^{\alpha(T)}$ and activated
dynamics $\xi(t,T)\propto {({\ell n}t)}^{1/\psi}$ compare well with simulation
fits, but possess too small a prefactor for activated dynamics.
\end{abstract}
\pacs{PACS numbers: 75.50.Lk, 75.10.Nr, 75.40.Gb} 
}
\narrowtext
The study of the irreversible behavior of the spin glass magnetization under a
change of magnetic field allows exploration of the available states of a
random
frustrated system.\cite{1,2} There are various representations for the long
time evolution and the dynamics of spin glasses,\cite{3,4,4a} but a coherent,
overall accepted real space description remains lacking.\cite{5}  The purpose
of this paper is to extract a time and temperature dependent spin glass
correlation length from a specially structured set of experiments, and to
compare our results with available theoretical predictions.

The definition of a correlation length for a spin glass is difficult to
express
in measurable terms.  Marinari et al. \cite{6} and Kisker et al. \cite{7}
introduced the time dependent equal time correlation function at time t.  In
the notation of Ref. 7,
$$G(x,t)={V^{-1}}\overline{{\sum
_i}<{\sigma_{i+x}}{\tau_{i+x}}{\sigma_i}{\tau_i}>_t}~~,\eqno(1)$$
where the average is done at time $t$, and ${\sigma_i}~({\sigma_{i+x}})$ and
${\tau_i}~({\tau_{i+x}})$ represent the $z$ component of Ising spins at sites
$i$ ($i+x$) in two thermalized configurations in a box of volume $V$.  To
avoid
accidental contributions to $G(x,t)$, the two configurations are chosen to
have
zero overlap,
\begin{math}
q={V^{-1}}{\sum_i}{\sigma_i}{\tau_i}.
\end{math}
Refs. 7 and 8 both observed, through their simulation studies, that for large
times $t$ the correlation function $G(x,t)$ differs from zero for distances
not
too much larger than a dynamic correlation length $\xi (t,~T)$.
Simulations of
Marinari et al.\cite{6} obtain satisfactory fits for $\xi(t,~T)\propto
(t/{\tau_0})^{\alpha T/{T_g}}$, appropriate to power law dynamics,\cite{4}
while Kisker et al. \cite{7} fit satisfactorily both this proportionality and
equally well $\xi(t,~T)\propto [(T/{T_g}){\ell n}({t}/{\tau _0})]^{1/\psi }$,
appropriate to activated dynamics.\cite{4a}

Our measurements consist of cooling a sample in a magnetic field through the
glass temperature $T_g$ to the measuring temperature $T$, waiting a time
$t_w$,
then cutting the field to zero and measuring the decay of the magnetization.
This generates the response function,
$$S(t)=d\Bigl[{-{M_{TRM}}(t,~{t_w})\over H}\Bigr]/{d\ell n\, t}~~,\eqno(2)$$
where ${M_{TRM}}(t,{t_w})$ is the thermoremanent magnetization at time $t$
after cutting the magnetic field to zero.

Our approach, justified previously through magnetic field cycling\cite{8} and
used to determine the Parisi physical order parameter $P(q)$,\cite{1} makes
use
of the scaling relationship introduced by Vincent {\it et al.}\cite{2}.  They
show that barrier heights surmounted during aging are reduced upon a change in
magnetic field by $E_z$, a quantity related to the change in Zeeman energy.
Our model\cite{1} assumes that barriers $\Delta~<{E_z}$ in the initial state
(before the magnetic field is changed) are quenched, with the population of
occupied states transitioning essentially instantaneously to those states of
lowest energy corresponding to the new value of the magnetic field.  The
associated change in magnetization is referred to as the reversible part of
the
magnetization.  The states ``left behind'' comprise the irreversible
component,
and decay by diffusion to the sink created by the quenching of barriers
$\Delta~<~{E_z}$.

At small magnetic field changes, states at the barrier height corresponding to
the waiting time $t_w$,
$$\Delta ({t_w},~T)={k_B}T(\ell n{t_w}-\ell n{\tau _0})~~,\eqno(3)$$
where 1/$\tau _0$ is an attempt frequency $\approx~{k_B}{T_g}/\hbar$, diffuse
towards the sink, resulting in a local equilibration of state occupancies on a
time scale of the order of $t_w$.  This causes\cite{9} a peak in the response
function $S(t)$, Eq. (2), for measurement times $t$ close to the waiting time
$t_w$.  When the magnetic field change is increased, $E_z$ increases, and the
{\it effective} barrier height from which diffusion takes place to the sink is
$\Delta({t_w},~T)-{E_z}$.  This results in a {\it reduction} of the measuring
time at which $S(t)$ peaks, and was first noted for experiments upon the
insulating thiospinel $Cd{Cr_{1.7}}{In_{0.3}}{S_4}$ by Vincent et al,\cite{2}
and the amorphous system ${({Fe_x}{Ni_{1-x}})_{75}} {P_{16}}{B_6}{Al_3}$ by C.
Djurberg et al.\cite{9a}  We follow their analysis and
associate the peak in $S(t)$ with an effective waiting
time,$t_w^{e\!f\!f}$, so
that as a function of magnetic field change,
$$\Delta({t_w},~T)-{E_z} ={k_B}T(\ell n{t_w^{e\!f\!f}}-\ell n{\tau
_0})~~.\eqno(4)$$

Systematic experiments were carried out by us for the metallic spin glass
$Cu:Mn~6at.\%$ and for the thiospinel.\cite{10}  Representative curves for
$-{M_{TRM}}/H$, $S(t)$ are exhibited in Fig. 1.

\begin{figure}
\epsfysize=3.7in \epsfbox{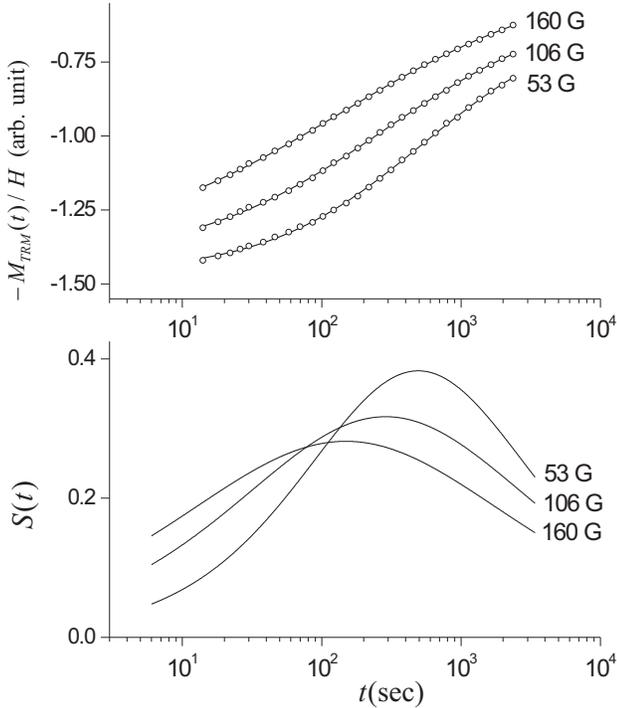}
\caption{(a)Plots of the data
points for $-{M_{TRM}}/H$ {\it vs} measurement time $t$, with the
solid lines the analytic fit; and (b)$S(t)$ defined in Eq. (2)
{\it vs} $t$ for $Cu:Mn~6at.\%$ at various changes in magnetic
field. $S(t)$ was calculated by taking the time derivative of the
analytic fit (solid line) in (a).} \label{fig1}
\end{figure}

Studies were performed for a variety of waiting times, $t_w$, magnetic field
changes, $H$, and temperatures, $T$, enabling us to extract values of $E_z$
over a wide range of parameter space.

We interpret $E_z$ to be the magnetic energy associated with a change in
magnetic field.  In the field cooled state, we associate each spin to
possess a
magnetic susceptibility $\chi _{fc}$ {\it per spin}, which we can calculate
from the total value of the field cooled magnetic susceptibility ${M_{fc}}/H$
by dividing by the total number of spins $N$.  Then the number of spins
participating in barrier quenching (and barrier hopping) can be derived from
setting,
$${E_z}={N_s}{\chi _{fc}}{H^2}~~.\eqno(5)$$
The quantity $N_s$ defines a volume over which the spins are effectively
locked
together for barrier hopping,\cite{2,10a} the radius of which we define as the
spin glass correlation length $\xi$.  Were $N_s$ to be the sum of smaller
independent clusters, the activation energy would relate to smaller barriers,
and $S(t)$ would not shift its peak from the vicinity of measurement times
$\approx~t_w$.
\begin{figure}
\epsfysize=3.7in \epsfbox{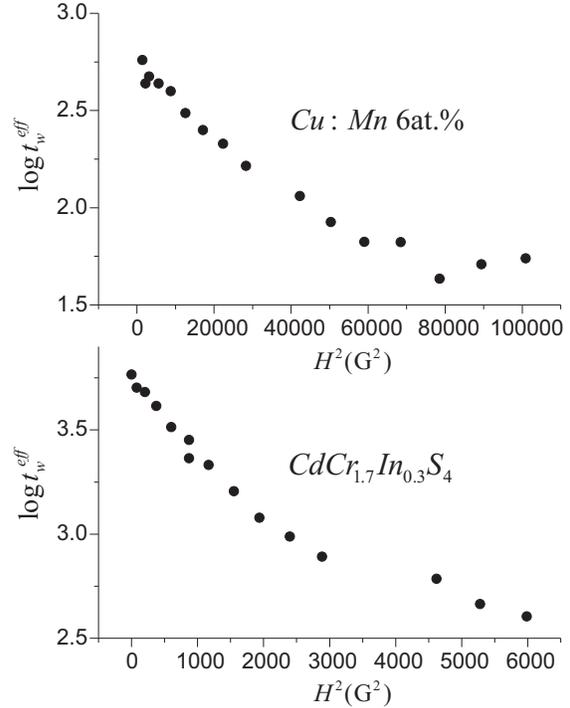}
\caption{A plot of $log~{t_w^{eff}}$ (equivalently, $E_z$ from Eq. 4) {\it vs}
$H^2$ for $Cu:Mn~6at.\%~({T/{T_g}}$ $=0.83,~{t_w}=480~sec)$ and
$Cd{Cr_{1.7}}{In_{0.3}}{S_4}~({T/{T_g}}=0.72,$ ${t_w}=3,410~sec)$ at fixed
$t_w$ and $T$.  The dependence is linear for magnetic fields less than $250G$
and $45G$, respectively, then breaks away to a slower dependence.  The
ratio of
these ``break'' fields is in agreement with the respective de Almeida-Thouless
lines.  We do not have a satisfactory explanation for this change in slope.  A
different description, Ref. 2, predicts a linear dependence of $E_z$ upon $H$,
which can be made to fit the data over the entire range of $H$ for the
thiospinel, but with a significant deviation at small field changes.  We have
chosen to focus in the text on the fit at small field changes because we want
to be certain to be in the linear regime for our analysis.}
\label{fig2}
\end{figure}
Extensive studies were made of both $Cu:Mn~6at.\%$ and
$Cd{Cr_{1.7}}{In_{0.3}}{S_4}$ to examine the $H^2$ dependence predicted in Eq.
(5). Experimental results for both systems, with $H$ scales adjusted to their
respective de Almeida-Thouless lines,\cite{10b} are exhibited in Fig. 2.

Experimentally, the measured quantity $N_s$ depends both upon waiting time
$t_w$ and temperature $T$.  It represents the number of correlated spin units
which flip together within a volume $\xi^3$ growing as a function of $t_w$ and
$T$.

We summarize our results for
$N_s$ in $Cu:Mn~6at.\%$ as a function of waiting time $t_w$ for fixed
temperature $T=0.89{T_g}=28K$ in Fig. 3.
The solid curve in Fig. 3 is a fit to the form derived from power law
dynamics,\cite{4}
$$\xi({t_w},~T)={0.653({t_w}/{\tau _0})}^{0.169{T/{T_g}}}~~,\eqno(6)$$
with $1/{\tau_0}=4.1\times {10^{12}}~{sec^{-1}}$.
A similar analysis for the thiospinel at $T=0.72{T_g}=12K$, but for only two
waiting times, $1,100~{\rm and}~3,410~sec$, gives
$\xi({t_w},~T)={0.53({t_w}/{\tau_0})}^{0.132{T/{T_g}}}$.  The two expressions
are remarkably close, and suggest that Eq. (6) may be universal (see Fig. 4
below).
\begin{figure}
\epsfysize=3.5in \epsfbox{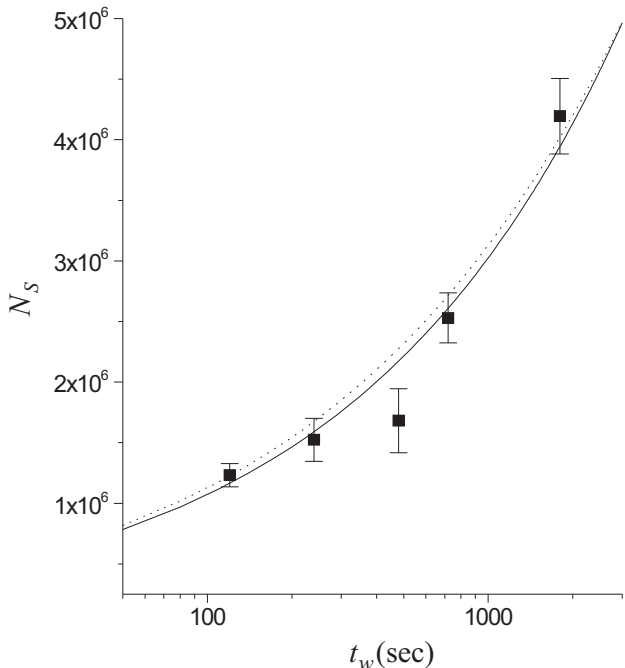}
\caption{A plot of $N_s$, the number of spins participating in barrier
quenching (and barrier hopping) from Eq. (4) for $Cu:Mn~6at.\%$, as a function
of Log ${t_w}$ at fixed temperature $T=0.89{T_g}=28K$.  The solid curve drawn
through the points is the prediction for power law dynamics, Ref. 4, while the
dashed curve is the prediction for activated dynamics, Ref. 5, with their
exchange factor set equal to $T_g$ (i.e. independent of $T$ and $t$).  As can
be seen from the two curves, the two fits are equally good.}
\label{fig3}
\end{figure}
\noindent
The dashed curve in Fig. 3 is a fit to the predictions of activated
dynamics,\cite{4a} leading to,
$$\xi({t_w},~T)={{10}^{-5}}{{[({T/{T_g}})\ell n({t_w}/{\tau
_0})]}^{1/0.21}}~~,\eqno(7)$$
where we have taken the nominally $t,~T$ dependent exchange factor of Ref.
5 to
be a constant equal to $T_g$.  As seen from Fig. 3, both Eqs. (6) and (7) fit
remarkably well.  However, the factors in Eq. (7) deserve comment.  The
prefactor is very troublesome, for it would give $\xi\approx O(1)$ for
${t_w}/{\tau_0}={10^5}$, while simulations\cite{6,7,10c} clearly see an
increasing dynamic correlation length in the regime $t/{\tau_0}<10^5$.  The
exponent in Eq. (7), equal to $1/\psi$, is also at the lower allowable
value,\cite{4a} $\psi=0.2$.

A comparison of Eq. (6) with the numerical results of the simulations in Refs.
\cite{6}, \cite{7}, and \cite{10c} is remarkable.  The prefactor in the first
is of order unity, with an exponent equal to $0.16T/{T_g}$ (compare with our
exponent $0.169T/{T_g}$), very close to our fitted results.  The second
finds a
prefactor of order unity and an exponent between $0.12T/{T_g}~{\rm
and}~0.13T/{T_g}$, and the third a prefactor close to unity, with an exponent
$0.13T/{T_g}$.  All three are in substantial numerical agreement with the
experimental result, Eq. (6).

\begin{figure}
\epsfysize=3.3in \epsfbox{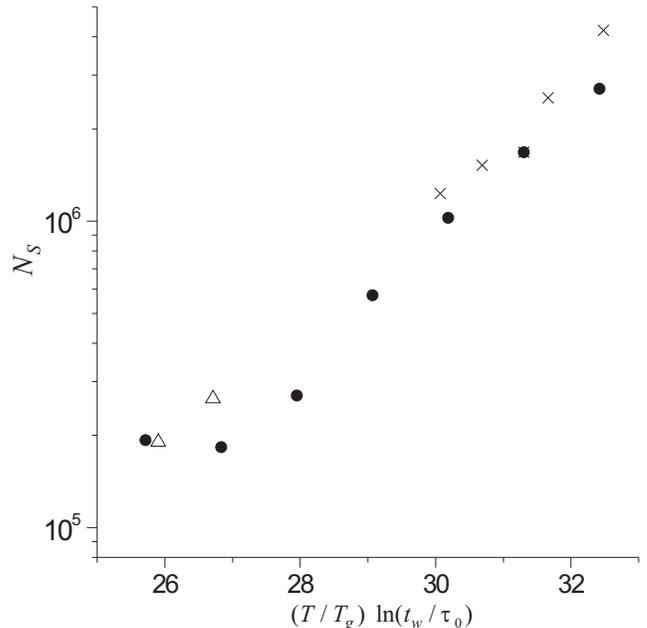}
\caption{A plot of $N_s$ on a log scale {\it vs} $(T/{T_g})\ell
n({t_w}/{\tau_0})$ for $Cu:Mn~6at.\%$ (solid circles and squares) and the
thiospinel (open triangles).  The solid circles are for variable $T$ and fixed
$t_w=8~min$, and the solid squares for variable $t_w$ and fixed
$T=0.89{T_g}=28K$ for $Cu:Mn~6At.\%$.  The open triangles are for variable
$t_w$ and fixed $T=0.72{T_g}=12K$ for the thiospinel.}
\label{fig4}
\end{figure}

In addition, Kisker et al.\cite{7} find a proportionality of the same form as
Eq. (7) for activated dynamics,\cite{4a} but with an exponent $1/\psi =
1/0.71\pm 0.02$, somewhat different from our fitted value $1/\psi=1/0.21$.
However, we have set the exchange factor of Ref. 5 equal to a constant,
$T_g$.
This quantity may, however, possess a ${t_w}$ and $T$ dependence over the
temperature region of our measurements.\cite{11}

In order to test the universality of our fit to the power law form for
$N_s$ as
given by Eq. (6), we plot in Fig. 4, $N_s$ on a log scale against
$({T/{T_g}}){\ln}({t_w}/{\tau_0})$, for both $Cu:Mn~6at.\%$ and the
thiospinel.  It is seen that all of our data falls roughly along a universal
line.  It should be noted that two data points for the thiospinel, taken at
low
temperatures [0.36 and 0.48 $\times T/{T_g}$, respectively], would not fall on
the data line in Fig. 4, and are not exhibited.  Both points correspond to
data
taken by a different method from those exhibited in Fig. 4, in that the
peak in
$S(t)$ could not be observed, and an estimate of the effective age of the
curves was obtained by a scaling procedure (see Fig. 5 of Ref. 2).  The
precise
fitting could be in error, but we cannot rule out that the ``line'' in Fig. 4
may crossover to a weaker slope at low temperatures.

The values of $N_s$ extracted by us from Eq. (4) are in the range of ${10}^5$,
varying with $t_w$ and $T$ according to Eqs. (6) and (7).  This suggests that
there is a mesoscopic number of spins participating as a unit in the aging
process, as argued initially in Ref. 2.  Lederman et al. and Hammann et
al.\cite{12} suggest a large highly correlated spin system is involved in the
dynamical processes of aging and irreversibility.

The nature of $G(x,t)$ in Eq. (1) lends itself to such an interpretation.  The
spin realizations $\sigma _i$ and $\tau _i$ have zero overlap.  Yet their
correlation is finite so long as $x<\xi(t,~T)$.  These realizations can be
interpreted as rotations one from another, consistent with a change in Zeeman
energy, relating the theoretical quantity $G(x,t)$ to $N_s$.

The power law dynamics form for $\xi({t_w},~T)$, exhibited in Eq. (6), and the
maximum occupied barrier $\Delta ({t_w},~T)$, stipulated by Eq. (3), generate
an interesting relationship.  Substituting $t_w$ from Eq. (6) into Eq. (3)
leads to,
$$\Delta ({t_w},~T)/{k_B}{T_g}=6[\ell n\xi({t_w}~,T)+0.44]~~.\eqno(8)$$
This result, supported by our experimental data, was anticipated by Koper and
Hilhorst\cite{4} as well as by Rieger\cite{10a} and Sibani {\it et
al.}\cite{13a}.  The form of Eq. (8) was also exhibited qualitatively in an
analysis similar to our own by P. Granberg et al. on
$Cu:Mn~10at.\%$.\cite{14}
Activated dynamics would yield a power law relationship between $\Delta
({t_w},~T)$ and $\xi({t_w},~T)$.

Equation (8) shows that it would be wrong for a finite time scale to think
of a
rigid set of barriers extending to infinite height (the so-called ``pure
states'').  Barriers are created as the correlation length increases, with
detailed balance obeyed throughout the hierarchy.

In summary, we have interpreted the magnetic field dependence of the response
function to generate the {\it number} of spins, ${N_s}({t_w},~T)$, locked
together in the barrier quenching (hopping) process.  We identify
${[{N_s}({t_w},~T)]^{1/3}}\approx {\xi}({t_w},~T)$ with the spin glass
correlation length.  The fitted expressions are nearly identical for
measurements on $Cu:Mn~6at.\%$, and for the thiospinel.  This suggests that
power law dynamics\cite{4} may well be universal.  Finally, our fitted
expressions for $\xi({t_w},~T)$ were very close to the numerical
simulations of
Refs. 7, 8, and 15.

The authors thank Drs. J.-Ph. Bouchaud, M. M\'ezard, H. Rieger, and P. Sibani
for very helpful conversations and correspondence.   We are especially
indebted
to Professor G. Parisi for pointing out the ${t_w}$ and $T$ dependence of
$\xi$
at an early stage of this work.  This work was supported by NSF Grant No. DMR
96 23195 and by the Japan Ministry of Education (Monbusho).


\begin{references}
\bibitem{1}
Y.G. Joh, R. Orbach. and M. Hammann, Phys. Rev. Lett. {\bf 77}, 4648 (1996);
Philos. Mag. B {\bf 77}, 231 (1998).
\bibitem{2}
E. Vincent, J.-P. Bouchaud, D.S. Dean, and J. Hammann, Phys. Rev. B {\bf 52},
1050 (1995).
\bibitem{3}
G. Parisi, Phys. lett. {\bf 73A}, 203 (1979); Phys. Rev. Lett. {\bf 43}, 1754
(1979); J. Phys. A {\bf 13}, L115 (1980).
\bibitem{4}
G.J.M. Koper and H.J. Hilhorst, J. Phys. France {\bf 49}, 429 (1988).
\bibitem{4a}
D.S. Fisher and D.A. Huse, Phys. Rev. B {\bf 38}, 373 (1988); {\bf 38}, 386
(1988).
\bibitem{5}
K. Jonason, E. Vincent, J. Hammann, J.-P. Bouchaud, and P. Nordblad, Phys.
Rev.
Lett. {\bf 81}, 3243 (1998).
\bibitem{6}
E. Marinari, G. Parisi, J. Ruiz-Lorenzo, and F. Ritort, Phys. Rev. Lett. {\bf
76}, 843 (1996).
\bibitem{7}
J. Kisker, L. Santen, M. Schreckenberg, and H. Rieger, Phys. Rev. B {\bf 53},
6418 (1996).
\bibitem{8}
D. Chu, G.G. Kenning, and R. Orbach, Philos. Mag. B {\bf 71}, 479 (1995); G.G.
Kenning, Y.G. Joh, D. Chu, and R. Orbach, Phys. Rev. B {\bf 52}, 3479 (1995).
\bibitem{9}
P. Nordblad, P. Svedlindh, J. Ferre, and M. Ayadi, J. Magn. Magn. Mater. {\bf
59}, 250 (1986); M. Ocio, M. Alba, and J. Hammann, J. Phys. (Paris) Lett. {\bf
46}, L1101 (1985); P. Granberg, L. Sandlund, P. Nordblad, P. Svedlindh, and L.
Lundgren, Phys. Rev. B{\bf 38}, 7097 (1988).
\bibitem{9a}
C. Djurberg, J. Mattsson, and P. Nordblad, Europhys. Lett. {\bf 29}, 163
(1995).
\bibitem{10}
In conjunction with V. Villar and V. Dupuis.
\bibitem{10a}
H. Rieger, J. Phys. A {\bf 26}, L615 (1993); H. Rieger, B. Steckemetz and M.
Schreckenberg, Europhys. Lett. {\bf 27}, 485 (1994).
\bibitem{10b}
J.R. de Almeida and D.J. Thouless, J. Phys. A{\bf 11}, 983 (1978).
\bibitem{10c}
J.-O. Andersson and P. Sibani, Physica A{\bf 229}, 259 (1996).
\bibitem{11}
J. Mattsson, T. Jonsson, P. Nordblad, H.A. Katori, and A. Ito, Phys. Rev.
Lett.
{\bf 74}, 4305 (1995).
\bibitem{12}
M. Lederman, R. Orbach, J. Hammann, M. Ocio, and E. Vincent, Phys. Rev. B {\bf
44}, 7403 (1991); J. Hammann, M. Ocio, R. Orbach, and E. Vincent, Physica
(Amsterdam) {\bf 185A}, 278 (1992).
\bibitem{13a}
P. Sibani, C. Sh\'on, P. Salamon, and J.-O. Andersson, Europhys. Lett. {\bf
22}, 479 (1993); P. Sibani and J.-O. Andersson, Physica A{\bf 206}, 1 (1994).
\bibitem{14}
P. Granberg, L. Sandlund, P. Nordblad, P. Svedlindh, and L. Lundgren, Phys.
Rev. B {\bf 38}, 7097 (1988).

\end{references}
\end{document}